\documentclass[conference]{IEEEtran}

\usepackage{xcolor}
\usepackage{lipsum}
\usepackage{siunitx}
\usepackage[ruled,vlined]{algorithm2e}
\usepackage{setspace}
\usepackage{subcaption}

\usepackage{cite}

\ifCLASSINFOpdf
  \usepackage[pdftex]{graphicx}
\else
    \usepackage[dvips]{graphicx}
\fi

\usepackage{amsmath}
\usepackage{algorithmic}

\usepackage{array}

\ifCLASSOPTIONcompsoc
\usepackage[caption=false,font=normalsize,labelfont=sf,textfont=sf]{subfig}
\else
\usepackage[caption=false,font=footnotesize]{subfig}
\fi

\usepackage{stfloats}
\usepackage{url}
\begin{document}

\title{Modeling Charging Demand and Quantifying Flexibility Bounds for Large-Scale BEV Fleets}

\author{\IEEEauthorblockN{María Parajeles Herrera\IEEEauthorrefmark{1}\IEEEauthorrefmark{2} and Gabriela Hug\IEEEauthorrefmark{1}}
\IEEEauthorblockA{\IEEEauthorrefmark{1}EEH - Power Systems Laboratory, ETH Zurich, Physikstrasse 3, 8006 Zurich, Switzerland}
\IEEEauthorblockA{\IEEEauthorrefmark{2}mparajele@ethz.ch}
}

\maketitle

\begin{abstract}
This paper presents a bottom-up method to model baseline charging power demand and quantify available flexibility for large-scale BEV fleets. The method utilizes geographic and sociodemographic information to represent the fleet's mobility and driving energy needs. It models the charging decisions of drivers based on their driving energy needs and range comfort level using real-world data. The flexibility quantification provides an hourly maximum and minimum bound for the charging power and limits the amount of daily flexible charging energy. We apply the methodology to the future fully electrified fleet of Switzerland as a case study and compare the spatio-temporal characteristics of the charging demand and flexibility of different geographic areas and urbanization levels. 
\end{abstract}


\IEEEpeerreviewmaketitle

\section{Introduction}

The charging demand of the future electrified passenger car sector poses challenges in the electric power system, as a new significant load with varying spatio-temporal charateristics must be supplied using, most often, non-dispatchable renewable generation sources. Understanding the charging demand patterns and their flexibility allows for better power system planning and operation since the charging demand impacts energy generation, transmission, and distribution. Therefore, modeling the future charging demand and quantifying its flexibility with high temporal and spatial resolution, from individual vehicles to national fleets, is necessary for a comprehensive planning and reliable operation of the future integrated electricity and transport sectors~\cite{Li2023}. 

Different approaches have been proposed to model charging demand. They have been primarily based on specific fleets' usage or charging networks' data to try and create representative charging profiles~\cite{Quiros-Tortos2018b, ZhangEV2023, Powell2022}. Nonetheless, these models are limited to the diversity of the data and are biased in best representing the characteristics of the drivers or the charging network considered, which reduces their representativeness of different regions. To overcome this, other studies have used mobility data as a starting point to model region-specific charging demand. For example, the authors in~\cite{Gschwendtner_flex_2023, Syla2024, Mangipinto2022} use travel survey information to first model the mobility patterns of EV drivers and derive charging energy needs. However, when calculating the charging demand, limiting assumptions like charging solely at home locations or during PV generation hours are used, which do not necessarily represent realistic charging behaviors. 

Quantifying the flexibility of the EV charging demand has been usually represented through a predefined set of charging profiles, which result from considering only a small number of possible charging behaviors or from freely shifting charging events within a time window, e.g., six hours~\cite{Gschwendtner_flex_2023, Mangipinto2022, Syla2024}. These sets aim to represent the shape change a charging profile can experience due to its flexibility. Nonetheless, the represented flexibility is limited to the charging behaviors considered, or overestimated due to the unrestricted shifting of charging events. Realistic quantification of the available charging flexibility has been less studied. The authors of~\cite{Muessel2023}, for example, propose a more realistic method of representing flexibility that quantifies the potential deviation from a base charging profile. Flexibility is quantified in terms of power but does not consider constraints on energy flexibility. This may overlook the users' minimum energy needs to complete subsequent trips or their preferences for the range available in their vehicles.

In this paper, we propose a charging demand model that captures realistic baseline charging behavior and a method to quantify the charging flexibility as a deviation from this baseline, considering both power and energy limitations. Together, they provide the basis to study the spatio-temporal characteristics of the charging demand and flexibility and the potential impacts and opportunities in their integration into the future power system. Hence, the main contributions of this work are: 

\begin{itemize}
    \item A bottom-up modeling method to represent mobility, driving energy, and user-comfort-based charging profiles.  
    \item A bottom-up charging flexibility quantification method that provides upper and lower power and daily flexible energy bounds. 
    \item A comparison of the spatio-temporal characteristics of charging demand and flexibility for a large-scale fleet of more than three million EVs driven in different geographic areas and urbanization levels (urban, periurban, rural).  
\end{itemize}

The remainder of this paper is structured as follows: Sections \ref{sec:drivingenergyneeds} and \ref{sec:charging_profiles} describe the proposed methods for modeling charging demand profiles, Sect. \ref{sec:charging_flexibility} explains the proposed method for flexibility quantification, Sect. \ref{sec:case_study} contains the case study results, and lastly, Sect. \ref{sec:conclusion} concludes the paper.  

\section{Geo-referenced driving energy needs}\label{sec:drivingenergyneeds}

This article builds upon the work developed in~\cite{ParajelesSEST2024}, where a framework was proposed for studying the spatio-temporal characteristics of driving energy needs starting from mobility patterns information.

The framework uses the output of the Multi-Agent Transport Simulation (MATSim), which provides georeferenced travel plans, i.e., transportation mode (e.g., passenger car) and the purpose (home, work, leisure, shop, or other errands), time, and duration of each 
trip for a statistically representative synthetic population~\cite{Horl2021}. The georeferenced mobility patterns are then processed and extended to a week-long dataset using open statistical data. A detailed traction, heating, and cooling driving energy needs estimation model is applied to every trip in the dataset, using geographic-specific weather information for four typical weather days in a year. We direct the reader to~\cite{ParajelesSEST2024} for more details of the mentioned framework and its results. The content of the output dataset of \cite{ParajelesSEST2024} is the basis for the methods described in the following sections. 


\section{Modeling baseline charging demand}\label{sec:charging_profiles}

To move from mobility patterns and driving energy needs to charging patterns, it is necessary to model the decision process of EV users to plug in (or not) and to start (or not) charging. The most important factors in this decision are the user's future driving energy needs (i.e., their mobility needs) and their vehicle's battery SOC comfort level~\cite{Gonzalez2021}. Hence, in this work, we model the plug-in-to-charge decision process based on the probability of drivers plugging in for each parking event depending on their vehicles' battery SOC~\cite{Pareschi2020}, and rule-based heuristics, which ensure a rational charging decision at all times.

\begin{figure}[ht]
\centering
\includegraphics[width = 1\linewidth]{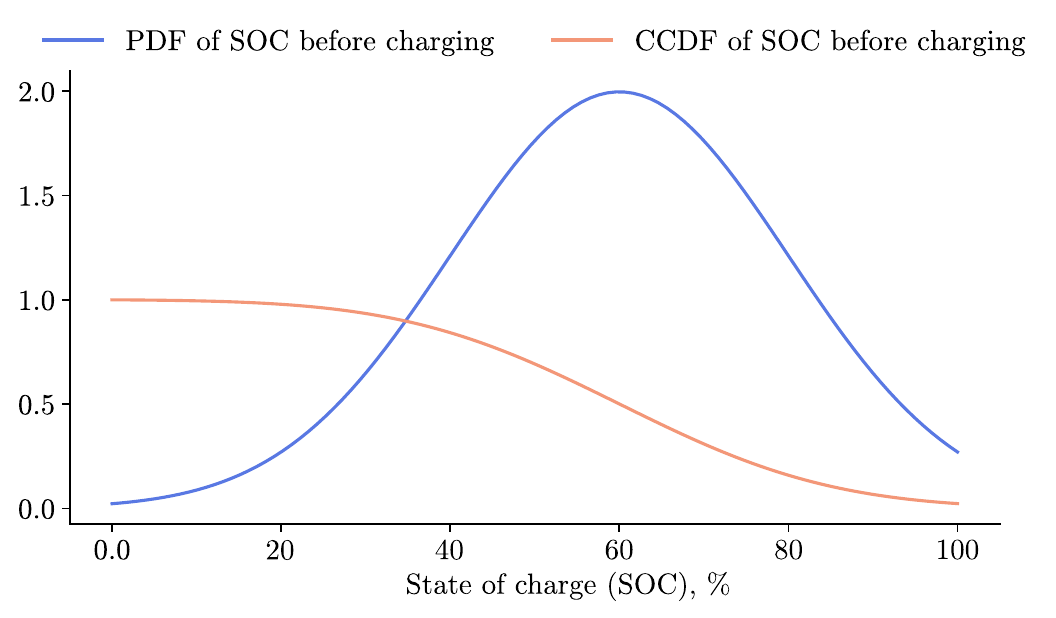}
\caption{Truncated PDF $\mathcal{T}(0.6, 0.2)$ and survival function for modelling charging decision~\cite{Pareschi2020}} \label{fig:chargingprobability}
\end{figure}

Algorithm \ref{alg:charging_profiles} describes in detail the model. For the battery SOC-related portion of the decision process, we use information from~\cite{Pareschi2020}, where the authors studied four different real-world EV trials charging data, which provide probability distribution functions (PDFs) that show the battery SOC distribution when a positive charging decision takes place. Fig.~\ref{fig:chargingprobability} shows the distribution found when considering the information from all trials. The PDF is a truncated normal distribution between $[0, \infty]$ (to ensure a positive charging decision for a fully depleted battery), with parameters $\mu\ =\ 0.6$, and $\sigma\ =\ 0.2$. The reverse of the cumulative distribution function (CCDF, or survival function) can be interpreted as the probability of a positive charging decision as a function of the battery's SOC at arrival at the charging location.  Notice that according to the PDF in Fig.~\ref{fig:chargingprobability}, most positive charging decisions take place at rather high battery SOCs. Even though the charging decisions also depend on the battery size and other factors such as electricity price and charger availability, recent studies have also found charging at high battery SOCs for different battery-sized fleets and locations in the US, Europe, and Australia~\cite{PhillipT2023, Powell2022}. 

\begin{algorithm}[ht]
\setstretch{1}
\SetAlgoLined
\caption{Charging decision and process.}
\label{alg:charging_profiles}
\SetKw{KwInput}{\textbf{Input:}}
\SetKw{KwOutput}{\textbf{Output:}} 
\KwInput{Mobility patterns and driving energy needs}\\
\KwOutput{Week-long charging profiles}

Set $SOC_{initial}$ = 100\% for every driver\\
\ForEach{day in 2 week-days and one week}{
    \ForEach{driver $d$ and parking event $p$}{
        \If{$t_p > 1$~h}{
            Get SOC$_{\text{sampled}} \sim \mathcal{T}(0.6, 0.2)$ \\
            \uIf{
            SOC$_{\text{arrival}}~<~$SOC$_{\text{sampled}}$ \textbf{or} \\
            \textcolor{white}{if: }SOC$_{\text{arrival}} < 15\%$ \textbf{or} \\
            \textcolor{white}{if: }SOC$_{\text{arrival}}$ insufficient for next two trips\\
            \textcolor{white}{if: }plus keeping 15\% SOC \textbf{or}\\
            \textcolor{white}{if: }remaining charging time insufficient to\\
            \textcolor{white}{if: }reach $SOC_{initial}$ 
            }{
                $charge\ decision \gets Positive$
            }
            \Else{
               $charge\ decision \gets Negative$
            }
            \If{$charge\ decision\ =\ True$}{
                Set $SOC_{final}\ \in \{80\%, 100\%\}$ where $P(SOC_{final} = 80)\ =\ p_{80}$ and $P(SOC_{final} = 100)\ =\ p_{100}$
                
                Charge during full $t_p$ or until $SOC_{final}$ 
            }
        }
    }
}
\end{algorithm}

To track the vehicles' battery SOC throughout their mobility patterns, we require information on their battery sizes. We choose a selection of batteries based on the distribution of sizes observed for the region of interest and match each vehicle's daily consumption requirements to one of the batteries available in the selection. For example, for Switzerland we consider batteries between \SI{70}{\kilo \watt \hour}, to \SI{120}{\kilo \watt \hour} in steps of \SI{10}{\kilo \watt \hour}~\cite{AutoSchweiz}. 

The charging decision process algorithm is applied to every parking event of all users during their week-long mobility and driving energy patterns described in Sect. \ref{sec:drivingenergyneeds}. Since there is no knowledge of the batteries' initial SOC, we perform a two-day initialization simulation by setting all vehicles batteries' SOC at 100\% at the beginning and using their final SOC (at the end of the two days) as starting SOC for the subsequent simulated week~\cite{Gschwendtner2023}. 

We assume that there is a widely available charging infrastructure since we want to capture people's charging patterns solely based on their energy needs and their vehicle's battery SOC level preferences (comfort level). The charging power rates considered are fixed but vary for each type of location, i.e., home $\SI{7}{\kilo \watt}$, workplace $\SI{11}{\kilo \watt}$ and public charging $\SI{22}{\kilo \watt}$~\cite{Gschwendtner_flex_2023}. 

The charging decision is made by every driver at every parking event by first checking if the parking event is long enough (e.g., at least one hour) to consider charging and then by comparing their vehicle's battery SOC at arrival to a sampled battery SOC from the PDF in Fig.~\ref{fig:chargingprobability}. The charging decision is positive if the SOC is lower than the sampled one. We include further constraints that help us model a rational charging behavior. We ensure that users do not drive with battery SOCs below 15\%~\cite{Quiros-Tortos2018b}; that their battery has enough energy to fulfill the following two trips; and that, by the end of the week, the vehicle ends with the same SOC with which it started. This reduces the computational burden of calculating the SOC across longer time scales (e.g., a month, a year) and allows for a weekly repetition of charging profiles without concerns of sudden SOC increases or decreases in each week's transition. The charging decision algorithm is applied to four typical weather weeks representing each year's season. 

The charging process is modeled by assuming the vehicle charges at the full charging rate possible and lasts until the parking time is up or until the final battery SOC is reached. The final battery SOC can be either 80\% or 100\% in an attempt to include in the model a higher level of charging literacy among users. It is set for each charging event based on the probability given in~\cite{Quiros-Tortos2018b}.  

Since all charging decisions and charging processes are modeled for each driver, we aggregate their charging patterns on an hourly basis to estimate the total charging demand of any region, e.g., defined by geographical and political boundaries, or, for instance, served by a particular distribution network. 

\section{Quantifying charging flexibility}\label{sec:charging_flexibility}

The flexibility calculation is based on the potential charging deviation from the baseline charging demand estimated using the method described in Sect. \ref{sec:charging_profiles}. This method of flexibility estimation captures both the inflexible portion of the charging strategy, i.e., users requiring a certain state of charge level after each charging event, and the potential deviation from the baseline charging strategy, which results in a lower chance of flexibility overestimation since it is limited by the user's mobility constraints~\cite{Muessel2023}. We consider unidirectional charging flexibility limited to shifting energy demand within a parking event and do not allow shifting between parking events. This is a conscious decision for our study in order to model the most straightforward flexibility to use, i.e., the one where the driver's comfort and decisions are minimally impacted. 

\begin{figure}[ht]
\centering
\includegraphics[width = 0.85\linewidth]{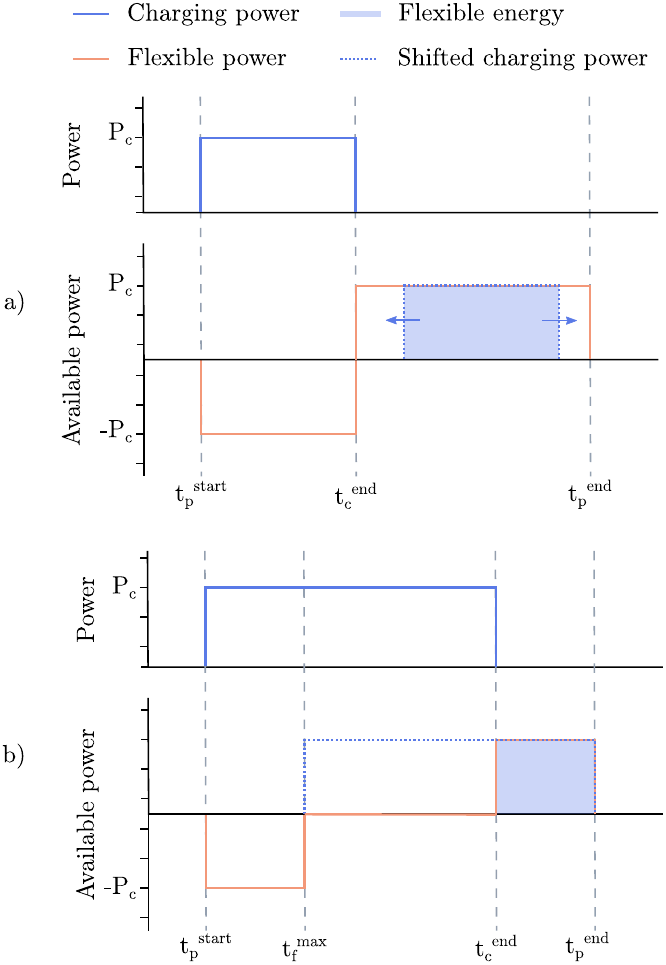}
\caption{Identification of flexibility bounds for two different charging events.}\label{fig:flexiboundsconcept}
\end{figure}

We estimate the charging flexibility for each driver and each charging event, continuing the bottom-up approach used to quantify the driving energy needs and model the baseline charging demand. The flexibility is quantified using three metrics: the charging power flexibility, expressed as 1) an hourly lower bound, which represents the inflexible part of the charging load; 2) an hourly upper bound, which represents the full charging power available resulting from also considering plugged-in vehicles in an idle state; and 3) the amount of energy that is considered flexible and can be shifted during a day which is used to bound the deviation from the baseline charging. This approach better reflects the reality in the flexibility quantification, even though we neglect the dependency of available flexibility on used flexibility, which is highly computationally and data expensive~\cite{YilinWen2023}. In Algorithm \ref{alg:flexibility_quantification}, we summarize the methodology proposed to quantify the charging flexibility and Fig.~\ref{fig:flexiboundsconcept} showcases graphically the calculation for two exemplary charging events. 

For the flexibility analysis, we first filter for only those charging events where flexible charging is realistic, i.e., where charging does not take the entire parking time and the parking time is long enough (e.g., at least one hour) but also not overly long, as vehicles plugged in for more than, e.g., fifteen hours, is unrealistic.

\begin{algorithm}[!b]
\setstretch{1} 
\SetAlgoLined
\caption{Flexibility quantification.}
\label{alg:flexibility_quantification}
\SetKw{KwInput}{\textbf{Input:}}
\SetKw{KwOutput}{\textbf{Output:}} 
\KwInput{Mobility patterns and charging profiles}\\
\KwOutput{Power and energy flexibility bounds}

\ForEach{driver, $d$, and parking event, $e$,}{
    \If{$e$ is a charging event \textbf{and} \\
     \textcolor{white}{if:}$t_p~\geq~1.05~\cdot~t_c$ \textbf{and} \\
     \textcolor{white}{if:}$1\text{h}~\leq~t_p~\leq~15\text{h}$}{
        \uIf{$t_p \geq 2 \cdot t_c$}{
            $E_f \gets E_{\text{ch}} \gets (t_{c}^{\text{end}} - t_{p}^{\text{start}}) \cdot P_{\text{ch}}$ \\
            $P_{\text{up}} \gets P_{\text{ch}}, \quad \forall \, t \in [t_{c}^{\text{end}}, t_{p}^{\text{end}}]$ \\
            $P_{\text{down}} \gets -P_{\text{ch}}, \quad \forall \, t \in [t_{p}^{\text{start}}, t_{c}^{\text{end}}]$ 
        }
        \Else{
            $t_{\text{f}}^{\text{max}} \gets t_{p}^{\text{end}} - t_{c}$\\
            $E_{\text{f}} \gets (t_{c}^{\text{end}} - t_{p}^{\text{end}}) \cdot P_{\text{ch}}$ \\
            $P_{\text{up}} \gets P_{\text{ch}}, \quad \forall \, t \in [t_{c}^{\text{end}}, t_{p}^{\text{end}}]$ \\
            $P_{\text{down}} \gets -P_{\text{ch}}, \quad \forall \, t \in [t_{p}^{\text{start}}, t_{\text{f}}^{\text{max}}]$ \\
            $P_{\text{up}} \gets P_{\text{down}} \gets 0, \quad \forall \, t \in [t_{\text{f}}^{\text{max}}, t_{c}^{\text{end}}]$ 
        }
    }
}
\ForEach{region $r$}{
    Aggregate $P_{\text{up}}$, $P_{\text{down}}$, and $E_f$ for all drivers in $r$
}
\end{algorithm}

We distinguish between two cases. First,  when there is complete freedom on when the charging event can take place because the parking time is at least twice as long as the charging time (Fig.~\ref{fig:flexiboundsconcept} a)), and thus all the charged energy is considered flexible. Second, when flexibility can be offered only during part of the charging event because charging takes most of the parking time (Fig.~\ref{fig:flexiboundsconcept} b)), and thus, only a portion of the charged energy can be considered flexible. 

As depicted in Fig.~\ref{fig:flexiboundsconcept}, we quantify the opportunities to lower the baseline charging power ($P_{down}$) or to charge in times where originally there was no baseline demand ($P_{up}$). The energy flexibility corresponds to either the full originally charged energy, as it can be charged in a time entirely outside the original schedule (Fig.~\ref{fig:flexiboundsconcept} a)), or to a portion of the originally charged energy, corresponding to the energy that can be charged outside the original schedule (as shown in Fig.~\ref{fig:flexiboundsconcept} b)). The difference between the original baseline charging profile and the calculated $P_{down}$ corresponds to the lower bound of charging power, which, if applied during the full considered time window, will not fulfill the required charging energy demand. Conversely, the sum of the original baseline charging profile and the calculated $P_{up}$ corresponds to the upper charging power bound. The flexible energy is computed as a daily value. For the overnight charging events, the flexible energy is added to the flexibility for both days. 

Since the flexible energy is quantified for each driver, we can quantify a regional lower and upper power flexibility bound and total daily flexible energy by aggregating all charging events occurring in a particular region. This formulation not only allows for comparing the spatio-temporal characteristics of different regions' demand and flexibility but also serves as a useful input for optimization frameworks, where optimized charging patterns are calculated using the power bounds and the daily flexible energy as constraints. 

\section{Case study} \label{sec:case_study}

In this section, we present the results for the charging profiles and the flexibility estimation when applying the methods described in Sect. \ref{sec:charging_profiles} and \ref{sec:charging_flexibility} to a future scenario with a fully electrified passenger-car fleet in Switzerland. Additionally, leveraging the bottom-up approach of the framework, we deep dive into a comparison of the results for smaller aggregations, i.e., municipalities, where we explore the nuances captured by the modeling framework related to the geographic location and the urbanization level of the regions under study. 

We use the mobility and parking patterns and the driving energy needs data from~\cite{ParajelesSEST2024}. It corresponds to georeferenced synthetic information of 3.2+ million BEVs, representing the mobile portion of the Swiss vehicle fleet in a day (ca. 70\%), and more than ten million trips~\cite{SFSOData2023}. The dataset contains the driving energy needs for each trip. The annual electric energy required for mobility is estimated at \SI{13.24}{TWh}, which represents approximately $22\%$ of the current national electricity demand~\cite{ParajelesSEST2024}. 

\begin{figure}[!b]
\centering
\includegraphics[width = 1\linewidth]{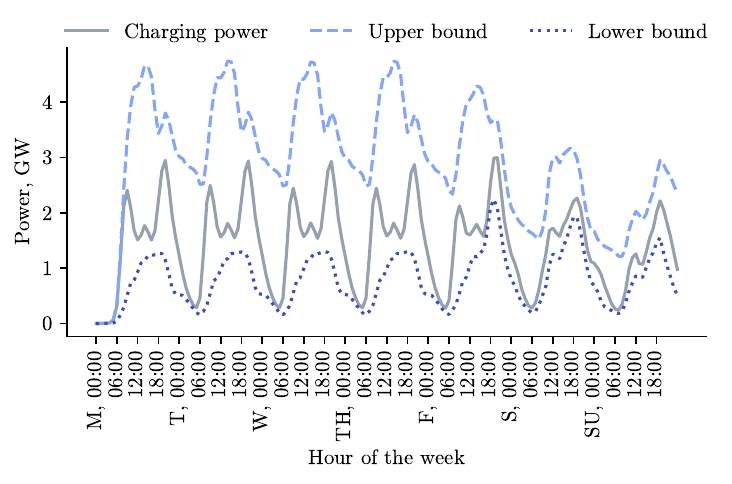}
\caption{Charging power and flexibility bounds in the coldest weather period.}\label{fig:chargingbounds}
\end{figure}

When studying the national-level results, we find the baseline charging profiles and lower and upper charging bounds as shown in Fig.~\ref{fig:chargingbounds}. The maximum peak corresponds to $\SI{3.07}{\giga \watt}$ during the weekdays and $\SI{2.3}{\giga \watt}$ during the weekend. The lower utilization of vehicles during the weekend also translates into lower charging flexibility, especially when also considering over-the-weekend constant plug-in status as unrealistic. The peaks are pronounced both in the early evening and at noon, marked by working hours patterns, as half of the workforce in Switzerland commutes by car~\cite{SFSOData2023}. 

\begin{figure}[!t]
\centering
\includegraphics[width = 1\linewidth]{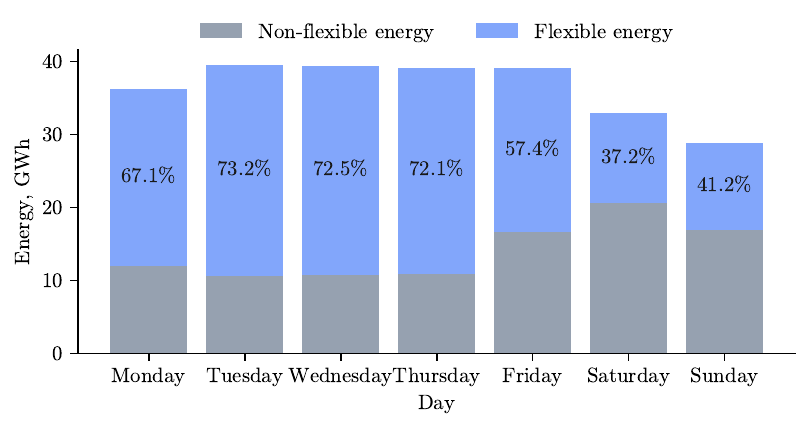}
\caption{Daily flexible energy, in the coldest weather period.}\label{fig:chargingflex}
\end{figure}

We found non-symmetrical lower and upper charging power bounds with respect to the baseline charging, with greater upward power flexibility. This is because shorter charging times are rather common since plug-in decisions on high SOCs occur frequently. We estimate an average of $\SI{1.2}{\giga \watt}$ of upward flexibility, with a maximum peak of around double in some hours of the day. The downwards flexibility is, on average, around $\SI{500}{\mega \watt}$, with a valley of around $\SI{1.2}{\giga \watt}$, which is significant given the overall Swiss generation capacity of around $\SI{27}{\giga \watt}$. 
Across the year, we also observed 16\% higher peaks in overall charging power on colder days with respect to warmer ones due to higher energy consumption in the former. Lastly, as shown in Fig.~\ref{fig:chargingflex}, we found an important difference in flexible energy during the week and weekends, with around $68\%$ of all charged energy being flexible on weekdays and about half on the weekends, reflecting reduced mobility and fewer flexible charging events, due to less parking at workplaces and longer parking periods at home without charging. 

\begin{figure}[!t]
    \centering
    
    \begin{subfigure}{\columnwidth} 
        \centering
        \includegraphics[width=0.89\columnwidth]{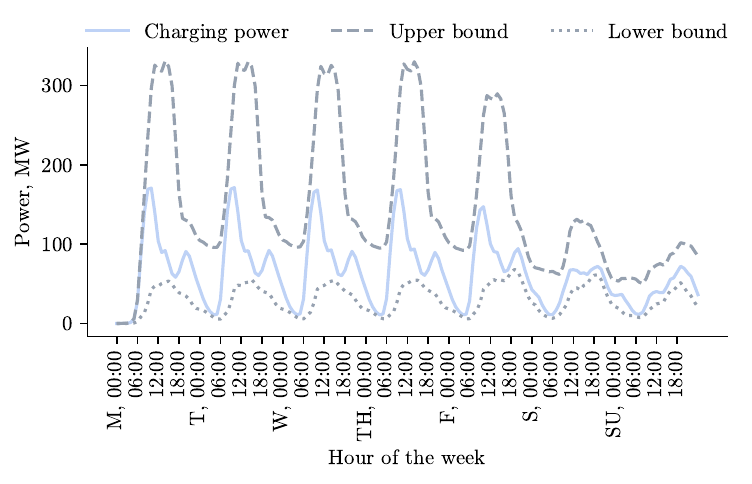} 
        \caption{Urban municipality: Zürich.}
        \label{fig:urban_zurich}
    \end{subfigure}
    
    \begin{subfigure}{\columnwidth}
        \centering
        \includegraphics[width=0.89\columnwidth]{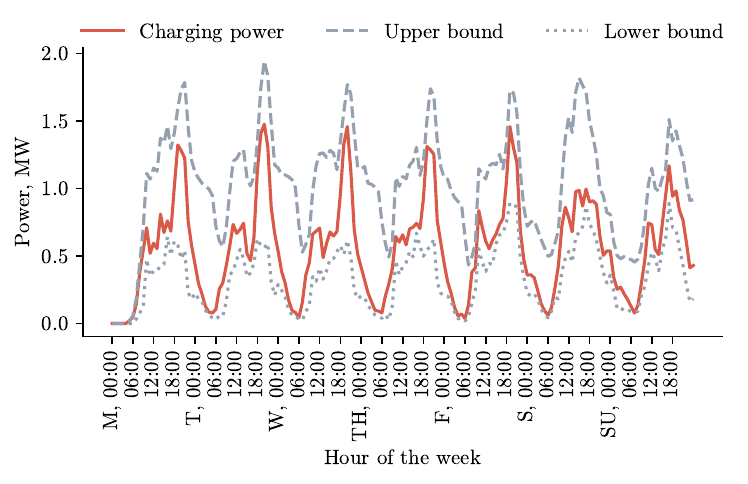}
        \caption{Rural municipality: Fischenthal.}
        \label{fig:rural_fischenthal}
    \end{subfigure}
    
    \begin{subfigure}{\columnwidth}
        \centering
        \includegraphics[width=0.89\columnwidth]{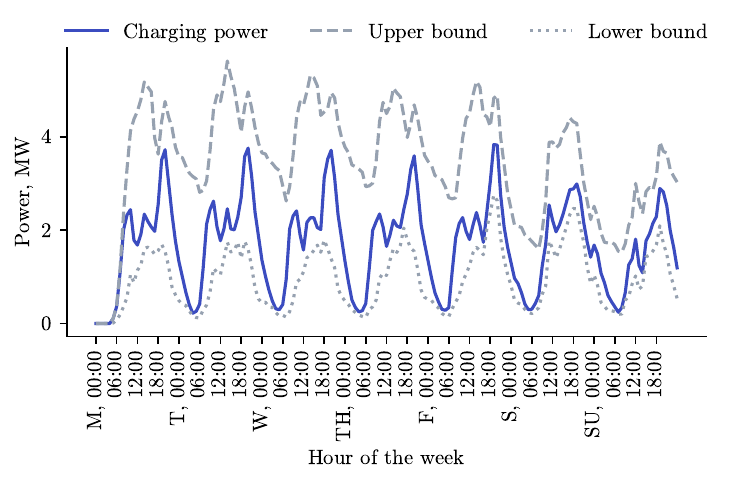}
        \caption{Periurban municipality: Pfäffikon.}
        \label{fig:periurban_pfaeffikon}
    \end{subfigure}

    \caption{Comparison of charging power and flexibility bounds for three municipalities, in the coldest weather period. }
    \label{fig:municipal_charging_comparison}
\end{figure}

Investigating the results at a municipality level, we compare three municipalities in the Swiss midlands with different urbanization levels in Fig.~\ref{fig:municipal_charging_comparison}. For the urban municipality, the city of Zürich, the model captures the high upward power availability during day-time hours, resulting from a majority of workplace- and public-based charging with high charging rates (i.e., \SI{11}{\kilo \watt}  to \SI{22}{\kilo \watt}), as depicted in Fig.~\ref{fig:urban_zurich}. A dominance of long and continuous parking events with shorter charging times (e.g., parking during work hours) also results in a stable upward power availability during daytime hours. Reduced home charging is evidenced both in a smaller evening peak and a lower upward and downward power availability during night hours. Fig.~\ref{fig:rural_fischenthal} shows the results for the rural municipality of Fischenthal. Most charging is home-based in rural areas, reflected by the clear evening-time charging peak. During night-time hours, upward charging flexibility is high, as a result of a high rate of plugged-in cars, but downward flexibility is limited, which reflects that most charging occurs through the night at slower charging rates. Week and weekend charging differ less than in urban areas since workplace charging is not dominant in the area. Downward flexibility is also more limited compared to urban areas, which results from higher driving energy needs per person due to longer distances in rural areas compared to city driving. Lastly, Fig.~\ref{fig:periurban_pfaeffikon} reflects the intermediate behavior of periurban zones, in this case, the municipality of Pfäffikon, where a combination of day-time charging (e.g., workplace- and public-based charging) at higher rates, and a majority of home-based charging at lower rates results in high upward power flexibility during the day.

\begin{figure}[htbp] 
    \centering
    \begin{subfigure}{\columnwidth}
        \centering
        \includegraphics[width=0.89\columnwidth]{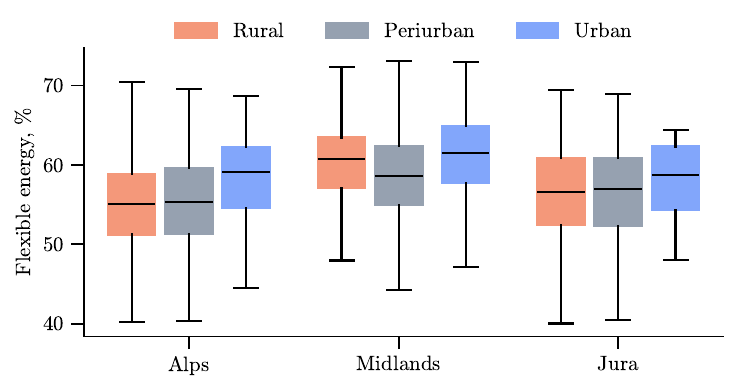}
        \caption{Numerical comparison of flexible energy.}
        \label{fig:Boxplot}
    \end{subfigure}
    \begin{subfigure}{\columnwidth}
        \centering
        \includegraphics[width=0.89\columnwidth]{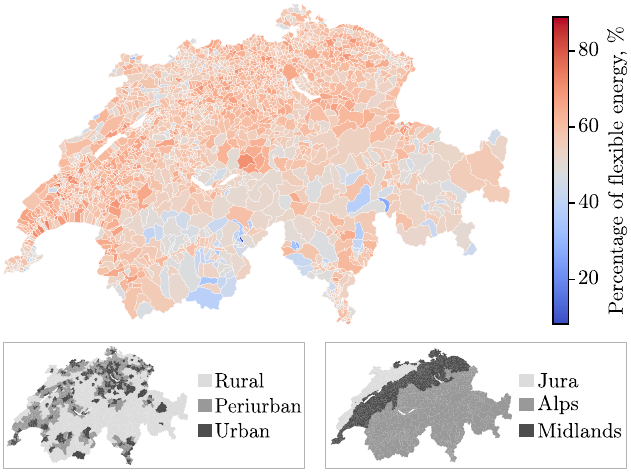}
        \caption{Geographic comparison of flexible energy. }
        \label{fig:Map}
    \end{subfigure}
    \caption{Flexible energy with respect to total charged energy in Swiss municipalities, in the coldest weather period.}
    \label{fig:flexpercentage}
\end{figure}

The percentage of weekly flexible charging energy for all considered municipalities across different geographic regions and urbanization levels are shown in Fig.~\ref{fig:Boxplot}. In all urbanization levels, between 55\% to 60\% of charged energy is flexible in most municipalities across the three geographic regions. Overall, the municipalities' flexible energy amounts vary between 40\% and 70\% of charged energy. Urban areas present approximately 5\% more flexible energy than others due to a higher idle time to charging time ratio. The spread of flexible charging energy percentages is highest in rural alpine areas, where the diversity of municipality types is also greatest. Geographically, the map in Fig.~\ref{fig:Map} shows a concentration of lower flexible energy percentages in alpine areas. In that region, driving energy needs per trip are higher than in the rest of the country due to longer distances and harsher weather and road conditions, which results in longer and more frequent charging requirements and, thus, less flexibility.  

\section{Conclusion}\label{sec:conclusion}
In this paper, we provide a method for modeling baseline charging profiles, quantifying charging flexibility, and studying the spatio-temporal characteristics of charging demand in diverse geographic locations and for different urbanization levels. In the case study, we found that the charging demand of the national-scale BEV fleet in Switzerland peaks both in the morning and at night, capturing the significant impact of both workplace and home charging, with the latter being the most prominent charging location. The weekday peak found corresponds to around 40\% of the current national peak, potentially reaching almost 60\% if using the full upward flexibility. The power bounds estimated are non-symmetrical due to fast charging in long parking events, which are the most common, especially in urban areas. The model also estimated charging demand during the week to be most flexible, reaching up to 70\% of charged energy in a day. In urban areas, we found that the most pronounced peak occurs during morning hours and that most charging occurs during the day and weekdays since workplace and public charging are most common, in contrast to rural areas where the most prominent peak occurs at night and the ratio between the weekend and week peak is lower. The proposed methodology captures the nuances expected in the areas explored. Its potential application varies from the evaluation of distribution grid loadability to generation capacity expansion planning studies.

\section*{Acknowledgment}

The research published in this publication was carried out with the support of the Swiss Federal Office of Energy (SFOE) as part of the SWEET EDGE project. The authors bear sole responsibility for the conclusions and the results presented.

\bibliographystyle{IEEEtran}
\bibliography{references}

\end{document}